\documentclass[preprint2]{aastex}

\slugcomment{To appear in Astrophysical Journal Letters}

\shortauthors{Vestergaard, Wilkes \& Barthel}
\shorttitle{Clues to QSO BLR Geometry and Kinematics}

\newcommand{\lya}{\ifmmode {\rm Ly\,}\alpha \, \else Ly\,$\alpha$\,\fi}
\newcommand{\hb}{H\,$\beta$}
\newcommand{\nv}{N\,{\sc v}}
\newcommand{\oiii}{[O\,{\sc iii}]}
\newcommand{\siivoiv}{Si\,{\sc iv}+O\,{\sc iv}]}
\newcommand{\civ}{C\,{\sc iv}}
\newcommand{\heii}{He\,{\sc ii}}
\newcommand{\feii}{Fe\,{\sc ii}}
\newcommand{\lam}{$\lambda$}
\newcommand{\kms}{\ifmmode {\rm km\,s}^{-1} \else km\,s$^{-1}$\fi}

\newcommand{\et}{\mbox{et~al.}\ }

\newcommand{\eg}{\mbox{e.~g.,}\ }

\newcommand{\hst}{{\it Hubble Space Telescope}}
\newcommand{\lsim}{\stackrel{\scriptscriptstyle <}{\scriptstyle {}_\sim}}
\newcommand{\gsim}{\stackrel{\scriptscriptstyle >}{\scriptstyle {}_\sim}}
\newcommand{\arcsecs}{\mbox{$^{\prime\prime}$}}
\newcommand{\logrv}{$\log R_{\rm V}$}
\newcommand{\mags}{\mbox{$^{m}$}}

\begin{document}

\title{Clues to Quasar Broad Line Region Geometry and Kinematics\footnote{
   Observations reported in this paper were obtained 
   at the Multi Mirror Telescope Observatory, a facility operated jointly 
   by the University of Arizona and the Smithsonian Institution.
   }}

\author{M. Vestergaard\altaffilmark{2,3,4}, B. J. Wilkes\altaffilmark{3}, 
and P. D. Barthel\altaffilmark{5}}

\altaffiltext{2}{Columbus Fellow. Present address: Dept.\ of Astronomy, 
The Ohio State University, 140 West 18th Avenue, Columbus, OH  43210-1173. 
Email: vester@astronomy.ohio-state.edu}
\altaffiltext{3}{Harvard-Smithsonian Center for Astrophysics, 60 Garden 
Street, MS-4, Cambridge, MA 02138. Email: bwilkes@cfa.harvard.edu.}
\altaffiltext{4}{Niels Bohr Institute for Astronomy, Physics and 
Geophysics, Copenhagen University Observatory, Juliane Maries Vej 30, 
2100 Copenhagen \O, Denmark.}
\altaffiltext{5}{Kapteyn Astronomical Institute, P.O.\ Box 800, NL-9700 
AV Groningen, The Netherlands. Email: pdb@astro.rug.nl. }


\begin{abstract}

We present evidence that the high-velocity \civ\,\lam 1549 emission line gas 
of radio-loud quasars may originate in a disk-like configuration, in 
close proximity to 
the accretion disk often assumed to emit the low-ionization lines.
For a sample of 36 radio-loud $z \approx$2 quasars 
we find the 20--30\% peak width 
to show significant inverse correlations with the fractional radio core-flux 
density, $R$, the radio axis inclination indicator.  
Highly inclined systems have broader line wings, consistent 
with a high-velocity field perpendicular to the radio axis.
By contrast, the narrow line-core shows no such relation with $R$, so the 
lowest velocity \civ -emitting gas has an inclination independent velocity 
field.  We propose that this low-velocity gas is located at higher 
disk-altitudes than the high-velocity 
gas.  A planar origin of the high-velocity \civ -emission is consistent with
the current results and with an accretion disk-wind emitting the broad lines.
A spherical distribution of randomly orbiting broad-line clouds and a polar 
high-ionization outflow are ruled out.

\end{abstract}



\section{Introduction \label{intro}}

Little is currently known about the detailed geometry and structure of the
broad-line region (BLR) in quasi-stellar objects 
(QSOs) due to the difficulty in spatially resolving the 
central regions even in the most nearby active 
galactic nuclei (AGN), the limited sensitivity of photo-ionization 
processes to the detailed BLR geometry, and to the complex nature of 
reverberation transfer functions (Peterson 1997; Netzer \& Peterson 1997).
For example, it is unclear whether the BLR gas has a spherical or 
disk-like distribution, and whether the BLR consists of discrete clouds or 
is part of a continuous gas distribution, such as an accretion disk wind 
[\eg Boroson \& Green 1992 (BG92); Murray \& Chiang 1997].
An improved knowledge of the geometry and structure of the centrally emitting 
and absorbing gas regions and their relation to one another and to the 
continuum source(s) may help us to better understand the observed 
continuum-line and line-line correlations (\eg BG92; Brotherton 1996; Wilkes 
\et 1999; Wills \et 1999), and furthermore shed light on the
interrelation of individual AGN subclasses.

The best constraints on the geometry and structure have so far been 
obtained from reverberation mapping (see the excellent reviews by 
Peterson 1993; Netzer \& Peterson 1997) of low-luminosity AGN showing
{(1)} the BLR is rather compact (the outer radius, R$_{\rm outer}$ is 
smaller than 0.3 pc $\approx 1\,\mu$arcsec), but is geometrically thick with 
R$_{\rm outer} \gsim$ 10 R$_{\rm inner}$,
{(2)} 
a certain degree of ionization stratification is present in the BLR gas, 
and
{(3)} the BLR size and the (bolometric) luminosity of the active nucleus 
are correlated (\eg Kaspi \et 2000).
Some clues to the relative position of the continuum source, and the
scattering, emitting, and absorbing components, respectively, can be
obtained from polarization studies of broad-absorption line QSOs 
(\eg Ogle 1997) which argue that they are highly inclined
``radio-quiet'' QSOs.
Additional and independent clues to the BLR gas kinematics may be obtained 
by studying the widths of the broad emission lines themselves. 
The gravitational potential of the central engine is thought to 
dominate the kinematics of the BLR gas and the line width (FWHM) to 
estimate the gravitational potential at the location of the gas, 
(\eg Peterson \& Wandel 1999).
If the motion of the BLR gas has a preferred direction, as 
opposed to disorganized virial motions around the central black hole,
we would expect the projection of this velocity field on our sight line to 
vary as the source axis inclination varies in the AGN population.

For a sample of broad-line radio sources 
Wills \& Browne (1986) found FWHM(\hb) to relate to the radio core-fraction, 
$\log R$, an estimator of the radio axis inclination 
to our line of sight, such that face-on sources lack very broad lines. 
They conclude that \hb\ is emitted by a disk configuration 
perpendicular to the radio axis, 
which offers 
support to the two-component model for the high and low ionization lines 
(Collin-Souffrin \et 1988).
The low-ionization, very high-density environment of the accretion disk fits 
the requirements of the physical conditions in the Balmer line emitting gas. 
The low level of long-term variability in the broad \hb\ wings (\eg Maoz \et 
1994; Peterson \et 2000) seems to support the disk interpretation 
[\eg Corbin \& Smith 2000; but see Brotherton (1996) for counter arguments].
Also, Zheng (1992) and Corbin (1997) model the high-velocity BLR emission 
finding consistency with Keplerian motion in low-$z$ sources.
Collin-Souffrin \et suggested a polar outflow origin for the high-ionization 
lines (HILs), thereby explaining their observed peak blueshift relative to the 
low-ionization lines (LILs; \eg Gaskell 1982; Wilkes 1984; Tytler \& Fan 1992). 
However, in the absence of a definitive model a spherical distribution of 
dense, discrete HIL-emitting clouds is often adopted in BLR studies to explain 
trends such as the inverse correlation between \oiii \,\lam\,5007 and the 
\feii\ emission (BG92).

In order to explore these possible models we studied the inclination 
dependence of the widths of \civ \lam 1549, representative of the HILs, 
relative to those reported for \hb\ in a sample of radio-powerful QSOs (RLQs).
We find interesting constraints applicable to the HIL geometry, which we 
present below. This project is part of an extended study of the ultraviolet 
(UV) broad emission lines in moderate-$z$ QSOs with a range of radio power.

\section{Data Analysis \label{dataproc}}

The details of the sample and the optical data 
of the extended study are described by Vestergaard \et (in 
preparation, hereafter paper~1).  We summarize the main points below.  
The radio data 
are described by Barthel \et (1988), Lonsdale, Barthel 
\& Miley (1993) and Barthel, Vestergaard \& Lonsdale (2000). 
The current sample consists of 25 lobe-dominated (LDQs; 
R$_{\rm 5} <$ 0.5; R$_{\rm 5}$  is defined below) and 11 core-dominated 
(CDQs; R$_{\rm 5} \geq$ 0.5) RLQs, chosen from 
the larger sample of $z \approx$ 2 RLQs, originally selected from Barthel, 
Tytler \& Thomson (1990), and the 3C and 4C catalogues. 
To ensure that we study the inclination dependence in the line widths, and 
not, say, a dependence on intrinsic source brightness, the CDQs and LDQs were 
selected such that their distributions of extended 5\,GHz radio luminosity, 
L$_{\rm ext}$, are similar.  A Kolmogorov-Smirnov test confirms this to a 
confidence level of 97.7\%.
The L$_{\rm ext}$-ranges and the mean and median values also match to within 
the standard error, while the M$_{\rm V}$-ranges match to within the expected 
uncertainty of $\sim$0.5$\mags$ (Table~1).
Three of the LDQs are compact ($\leq$15\,kpc) but their $R_{\rm 5}$-values 
could be measured from our high-resolution radio images.  Highly variable RLQs 
are excluded.
For the RLQs not presented by Barthel \et (1990) we obtained MMT Blue 
Channel spectra during 1994 October -- 1996 November.  The RLQs were 
observed at airmasses below 1.4 and with a 1\arcsecs\ slit placed at the 
parallactic angle.  The Barthel \et data were not well flux-calibrated and 
so were recalibrated using short MMT exposures.

Spurious noise spikes and narrow absorption lines superposed on the 
emission lines were corrected by generating smooth fits to the 
profiles.  The multiple Gaussian fitting (Laor \et 1994; paper~1) 
that was applied to the \civ --\heii\ line complex also allows a 
deblending of the line emission contributions, especially in the 
red wing of \civ, which in a few objects shows signs of weak \feii\ 
emission.  This is not expected to affect the current results adversely.
The limitations imposed by the \feii\ emission will be discussed in paper~1.
The profile parameters were measured on the smooth \civ\ fits to ensure that 
they are measured consistently, objectively, and with minimum uncertainties.
The full line widths at fractional profile heights,
Y (= 10\%, 20\%, ..., 90\%) were determined (FWYM), along with 
the interpercentile velocity widths, IPV(Y\%) (Whittle 1985). 
Reliable error estimates of each line parameter, based on continuum and flux 
uncertainties, were determined in a similar fashion as outlined by Whittle 
(1985; see paper~1 for details). 

The radio core-fraction is defined here as $R_{\rm 5}$ = S$_{\rm core}$/S$_{\rm 
total}$, where S$_{\nu}$$\sim$$\nu^{\alpha}$ are the observed fluxes at 5\,GHz 
measured from our VLA images; the spectral index, $\alpha$, is defined as
$\alpha^{\rm 408MHz}_{\rm 5000MHz}$ (or $\alpha^{\rm 1400MHz}_{\rm 5000MHz}$ 
if the former is unavailable). 
A K-correction\footnote{H$_0$ = 50 km~s$^{-1}$~Mpc$^{-1}$, q$_0$ = 0 is used 
throughout} was applied to $R_{\rm 5}$ to permit a comparison in the 
rest-frame: 
$R_{5} = S_{\rm core}/[(S_{\rm core}\,+\,S_{\rm ext.})(1+z)^{(\alpha_{\rm ext.} 
- \alpha_{\rm core})}]$, 
where $\alpha_{\rm core} = $0 and $\alpha_{\rm extended} =-$0.8 (\eg Kellerman 
\& Pauliny-Toth 1981).
Our definition of $R_{\rm 5}$ is less sensitive to variations 
in the density of the environment of the radio source than \eg
the $R$ defined as $S_{\rm core}/S_{\rm ext.}$, especially for face-on 
objects, where $S_{\rm ext.}$ tends to be relatively weak.  
We also consider the inclination measure, \logrv, defined as
the ratio of the 5\,GHz core flux density to the visual flux density with 
applied K-correction  (Wills \& Brotherton 1995). 
It is also less sensitive to the density of the radio source environment.

\section{Trends with Source Inclination}

FWHM(\civ) shows no relation with $\log R_{5}$ (the confidence 
level\footnote{Confidence level = 1 $-$ P, where P is the probability of 
no correlation} [herafter CL] of the Spearman rank correlation test is 
$\sim$79\%). 
However, FW20M(\civ ), representative of the width of the broad
\civ\ base, shows a strong inverse correlation with $\log R_{5}$ at the 
99.7\% CL (Fig.~\ref{fig1.fig}a). Support for the reality of this effect 
comes from the similar relations for FW10M, FW30M, and FW40M (CLs: 97.5\%, 
99.6\%, and 98.6\%).  
As an ``area parameter'', IPV(Y) is much less sensitive to the varying 
line-core strength which causes scatter in the FWYM measures (paper~1). 
The fact that IPV(40\,--\,70\%) show an inverse correlation with 
$\log R_{5}$ at the 97--98\% CL, also supports the reality of the FW20M trend.
The IPV(Y$<$40\%) is more sensitive to variations in the extreme wings
than FWYM, evident from the larger scatter and errors.
This does not jeopardize the above results for IPV(40\,--\,70\%) [paper~1].
The \logrv\ measure shows similar relations as $\log R_{5}$ though with 
relatively lower significance. This 
is most likely due to the fact that a {\it zone of avoidance} is more
evident in these relations (\eg Figure~1b) though it is less apparent than
that presented for \hb\ by Wills \& Browne (1986). 
Brotherton (1996), however, finds clear anti-correlations for FW25M and FWHM
of \hb\ (but none for FW75M).  

The \civ\ FW20-30M relations with core-dominance must primarily be driven by 
inclination and not some other property because: {\bf (1)} the L$_{\rm ext}$ 
distributions are similar for LDQs and CDQs. This is further confirmed by the 
at most weak FW20-30M -- L$_{\rm ext}$ relations (CL $\lsim$95\%), {\bf (2)} 
the lack of EW correlations with core-dominance (CLs $<$92\%) indicates that 
the (luminosity normalized) line flux does not change with inclination, 
{\bf (3)} the lack of FW70-90M relations with core-dominance (CLs $<$37\%) 
shows there is no weakening of the narrow line core in LDQs which could yield 
an apparent but unreal FW20-30M inclination dependence, and {\bf (4)} no 
correlation with inclination is found for M$_{\rm V}$, L$_{\rm ext}$, and 
redshift.

The line width relations with inclination directly show 
that the high-velocity field projected along the line of sight increases with 
source inclination.  The velocity field of BLR gas (in discrete clouds or a 
continuous distribution) in organized orbits around the source axis is expected 
to behave this way.  The similar \hb\ and \civ\ relations imply that, to first 
order, the kinematical structures of the two high-velocity line-emitting 
regions, and by inference those 
of HILs and LILs, are similar. Item (2) above also implies that the line and
continuum fluxes are anisotropically emitted to about the same degree.  Highly
beamed sources are excluded and the UV continuum is also not expected to be
significantly beamed (\eg Wills \& Brotherton 1995). Item (2) confirms this.

Though these results do not provide independent evidence for a disk-like 
geometry, they are consistent with the high-velocity HIL gas being physically 
close to the accretion disk and with its predominant motion being in a plane 
parallel to the disk. 
However, the \civ\ emission cannot originate in the low-ionization accretion 
disk-material itself, while the \hb\ emission may (\eg Collin-Souffrin \et 
1988).  We therefore propose that the high-velocity HIL gas 
is planar and lies in close proximity to the disk, and with similar overall 
kinematics. 
In our ongoing work we will re-address this FW20-30M inclination dependence
issue with larger and optimally constructed L$_{\rm ext}$-samples of LDQs and 
CDQs.

Our results argue against two possible BLR geometries: 
(i) the velocity field of a spherical distribution of randomly orbiting clouds 
is not expected to vary with inclination, and (ii) a polar 
outflow of the HIL gas would show the opposite inclination dependence to 
that seen here for \civ. 
However, a planar HIL velocity-field and geometry is consistent with a 
disk-wind, where the broad lines are emitted at the base of dense, optically 
thick winds evaporating off the accretion disk (Murray \& Chiang 1997, 1998). 
This model can successfully explain a number of observational details.
Bottorff \et (1997) and K\"onigl \& Kartje (1994) present alternative 
disk-like BLR models.

Item (3) above implies that the lowest-velocity BLR gas has a random velocity 
field relative to inclination.  It may not be physically associated with the 
high-velocity gas located close to the accretion disk. Low-density HIL gas of 
lower optical depth at higher disk-altitudes, similar to the proposed 
intermediate line region (ILR; \eg Brotherton \et 1994), fits this requirement; 
the altitude is low enough to maintain the high ionization level.
In the Murray \& Chiang (1997) disk-wind model the narrow line-core is emitted 
by the wind at much larger disk-radii than the broad line-base. 
This predicts similar inclination dependences for the high and low velocity
fields in contrast to our results and so requires a modification 
of the model, possibly along the lines suggested above.

Brotherton (1996) finds very similar results for the inclination dependence of 
the high and low velocity \hb\ emission in a sample of RLQs. 
The VBLR/ILR picture, which he advocates, is not geometry specific, 
but is fully consistent with the results and interpretations presented here.

A disk-like geometry of the HIL gas predicts:
{\bf (1)} other HILs (\lya, \nv, \siivoiv, and \heii) should show 
similar dependences on source inclination; such a study is underway,
{\bf (2)} the low-velocity dispersion ($\lsim$ 2000 \kms) of the narrowest 
line core of the HILs should be independent of inclination, 
assuming this gas is located at relatively high disk-altitudes, and
{\bf (3)} the \civ\ and \hb\ profile parameters should behave similarly 
with varying source inclination, excluding the \civ\ narrow line core and the 
contribution to \hb\ from the narrow line region.
No detailed study exists of high-redshift QSOs; the FWHMs of \civ\ and \hb\ 
correlate somewhat at low redshift (\eg Wilkes \et 1999). 
We are currently studying this issue for low-redshift QSOs, based in part 
on \hst\ archival data. 

\section{Summary \label{summary}}

Despite years of study little is known for certain about the 
geometry of the BLR of AGN beyond its compact size ($\lsim$light-weeks).
The present results indicate that the high-velocity HIL gas is located in 
a disk-like configuration close to the accretion disk, long thought to
emit the LILs, and with similar kinematics.
Our results argue against a spherical cloud distribution with random cloud 
velocities and a high-ionization polar outflow.
If confirmed, the implications are intriguing; they may help to constrain 
interpretations of BLR data and to suggest new studies to shed light
on the detailed structure and kinematics of the AGN BLR.



%
%

\acknowledgments

\vskip -0.2cm 
We thank Drs.\ Smita Mathur and Brad Peterson for comments on this paper, Dr.\ 
Bev Wills for useful discussions of the \logrv\ measure, and Dr.\ Norm Murray 
of the disk-wind model.
MV is very pleased to thank the Smithsonian Astrophysical Observatory for their 
hospitality and gratefully acknowledges financial support from the Columbus 
Fellowship at The Ohio State University, the Danish Natural Sciences Research 
Council (SNF-9300575), the Danish Research Academy (DFA-S930201) and a Research 
Assistantship at Smithsonian Astrophysical Observatory through NASA grants 
NAGW-4266, NAGW-3134, and NAG5-4089 to BJW.  BJW gratefully acknowledges 
financial support from NASA contract NAS 8-39073 ({\it Chandra} X-ray Center).
PDB acknowledges contributions by Drs.\ David Tytler and George Miley
in the early stages of this project, and travel support by the Leids
Kerkhoven-Bosscha Fonds.

\clearpage
\begin{deluxetable}{crrrrrr}
\tabletypesize{\normalsize}
\tablecaption{Subsample Match Statistics. \label{tbl-1}}
\tablecolumns{7}
\tablewidth{0pt}
\tablehead{
&& \multicolumn{4}{c}{$\log \rm L_{\rm ext}$ (5\,GHz)} & 
  \colhead{$\rm M_{\rm V}$} \\
\cline{3-6} 
\colhead{Sample} & \colhead{\#} & \colhead{Mean}   & \colhead{Median} & 
\colhead{Std.err.} & \colhead{Range}  & \colhead{Range}
}
\startdata

LDQ & 26 & 27.75 & 27.75 & 26.84 & 26.89 -- 28.13 &  $-$26.5 -- $-$28.9 \\
CDQ & 11 & 27.78 & 27.82 & 27.12 & 27.10 -- 28.11 &  $-$27.0 -- $-$29.2 \\
 \enddata


\end{deluxetable}

\clearpage

\clearpage

\figcaption[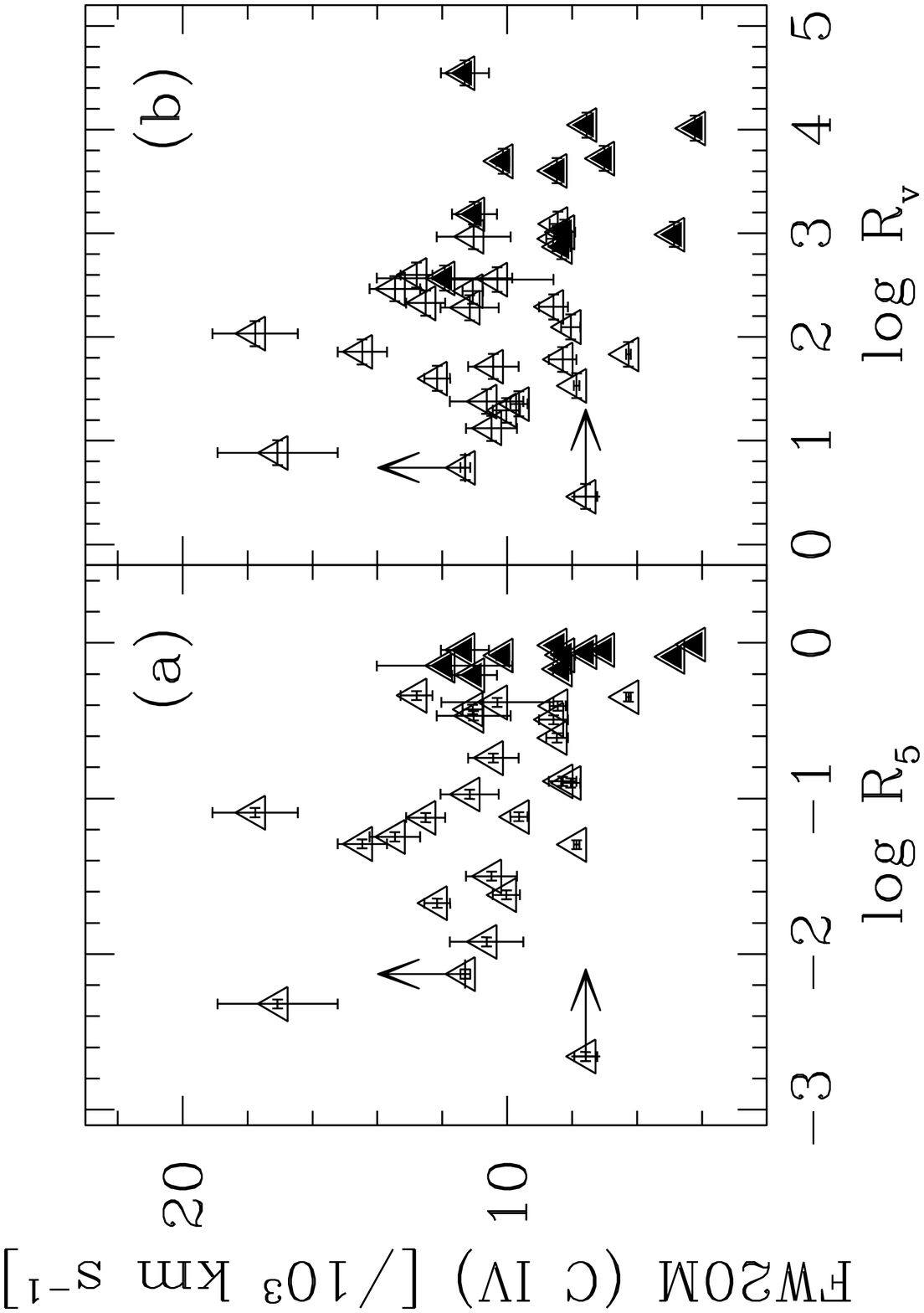]{The FW20M relations with (a) $\log R_{\rm 5}$ and (b) 
\logrv .  Core-dominated RLQs (CDQs) have solid symbols.
Spearman rank correlation tests show confidence levels of (a) 99.7\% and
(b) 96\%. The lower confidence levels for \logrv\ is likely due to
its relation being a {\it zone of avoidance} of very broad lines in CDQs
and not a clear inverse correlation.
 \label{fig1.fig}}

\clearpage
\vskip 3cm

\epsscale{2.0}
\plotone{fig1.eps}

\end{document}